\documentclass[12pt,notitlepage,fleqn]{article}%
\usepackage{amssymb}
\usepackage{graphicx}
\usepackage{amsmath}
\usepackage{indentfirst}%
\usepackage{amsfonts}
\begin{document}

\begin{center}
{\Huge A Proposal About}

{\Huge the Confinement of Quarks }

\bigskip

{\normalsize Jiao Lin Xu}

{\small The Center for Simulational Physics, The Department of Physics and
Astronomy}

{\small University of Georgia, Athens, GA 30602, USA}

E- mail: {\small \ Jxu@Hal.Physast.uga.edu}

\bigskip

\textbf{Abstract}
\end{center}

{\small From the Dirac sea concept and the Quark Model, we infer a
phenomenological accompanying excitation concept. This concept will give an
explanation for the confinement of quarks.}

\subsection{Introduction\ \ \ \ \ \ \ \ \ \ \ \ \ \ \ \ \ \ \ \ \ \ \ \ \ \ \ \ \ \ \ \ \ \ \ \ \ \ \ \ \ \ \ \ \ \ \ \ \ \ \ \ \ \ \ \ \ \ \ \ \ \ \ \ \ \ \ \ \ \ \ \ \ \ \ \ \ \ \ \ \ \ \ \ \ \ \ \ \ \ \ \ \ \ \ \ \ \ \ \ \ \ \ \ \ \ \ \ \ \ \ \ \ \ \ \ \ \ \ \ \ \ \ \ \ \ \ \ \ \ \ \ \ \ \ \ \ \ \ \ \ \ \ \ \ \ \ \ \ \ \ \ \ \ \ \ \ \ \ \ \ \ \ \ \ \ \ \ \ \ \ \ \ \ \ \ \ \ \ \ \ \ \ \ \ \ \ \ \ \ \ \ \ \ \ \ \ \ \ \ \ \ \ \ \ \ \ \ \ \ \ \ \ \ \ \ \ \ \ \ \ \ \ \ \ \ \ \ \ \ \ \ \ \ \ \ \ \ \ \ \ \ \ \ \ \ \ \ \ \ \ \ \ \ \ \ \ \ \ \ \ \ \ \ \ \ \ \ \ \ \ \ \ \ \ \ \ \ \ \ \ \ \ \ \ \ \ \ \ \ \ \ \qquad
\ \ \ \ \ \ \ \ \ \ \ \ \ \ \ \ \ \ \ \ \ \ \ \ \ \ \ \ \ \ \ \ \ \ \ \ \ \ \ \ \ \ \ \ \ \ \ \ \ \ \ \ \ \ \ \ \ \ \ \ \ \ \ \ \ \ }%

The Standard Model \cite{Standard} has been enormously successful in
explaining and predicting a wide range of phenomena. There is a weak point,
however, a rigorous basis for confinement of the quarks is needed
\cite{Confine}. M. Kaku has pointed out \cite{Kaku}:

\begin{quote}
{\small \textquotedblleft Although quark confinement has not been rigorously
proved within the framework of QCD, we provide compelling reasons for
believing that quarks are confined in lattice gauge theory. }

\qquad{\small In general, if the potential between two quarks is proportional
to the distance between them, then the two quarks can never be separated:\ }
\begin{equation}
\text{{\small Confinement: V(r)}}\sim\sigma\text{r}\label{Vr}%
\end{equation}
{\small \ where }$\sigma${\small \ is called the string tension. If we try to
separate the quarks by force, then the restoring force of the linear potential
between them grows sufficiently rapidly to prevent them from being separated.
Furthermore the string may break, creating a quark-antiquark pair held
together by another string. Thus, they can never be separated if they are
bound by a linear potential. Similarly, if the quark potential asymptotically
becomes a constant or decreases with distance, then the potential is not
sufficient to confine the quarks.\textquotedblright\ \ \ }
\end{quote}

{\small \ }This is a good idea. The question is: How do the two quarks form a
linear potential that increases with distance (not decreases)?
Since\ gravitational force and electric force all decrease with distance, if
there is a unitary form of the three forces, the strong potential between two
quarks will decrease with the distance also, so that the potential is not
sufficient to confine the quarks.

Using the potentials between quarks inside hadrons, most physicists try to
explain the confinement, forgetting the vacuum quark seas outside the hadrons.
From the vacuum quark seas concept, we infer a phenomenological
\textquotedblleft accompanying excitation concept\textquotedblright\ to try to
explain the confinement of the quark.\ \ 

\subsection{Quark Seas in the
Vacuum\ \ \ \ \ \ \ \ \ \ \ \ \ \ \ \ \ \ \ \ \ \ \ \ \ \ \ \ \ \ \ \ \ \ \ \ \ \ \ \ \ \ \ \ \ \ \ \ \ \ \ \ \ \ \ \ \ \ \ \ \ \ \ \ \ \ \ \ \ \ \ \ \ \ \ \ \ \ \ \ \ \ \ \ \ \ \ \ \ \ \ \ \ \ \ \ \ \ \ \ \ \ \ \ \ \ \ \ \ \ \ \ \ \ \ \ \ \ \ \ \ \ \ \ \ \ \ \ \ \ \ \ \ \ \ \ \ \ \ \ \ \ \ \ \ \ \ \ \ \ \ \ \ \ \ \ \ \ \ \ \ \ \ \ \ \ \ \ \ \ \ \ \ \ \ \ \ \ \ \ \ \ \ \ \ \ \ \ \ \ \ \ \ \ \ \ \ \ \ \ \ \ \ \ \ \ \ \ \ \ \ \ \ \ \ \ \ \ \ \ \ \ \ \ \ \ \ \ \ \ \ \ \ \ \ \ \ \ \ \ \ \ \ \ \ \ \ \ \ \ \ \ \ \ \ \ \ \ \ \ \ \ \ \ \ \ \ \ \ \ \ \ \ \ \ \ \ \ \ \ \ \ \ \ \ \ \ \ \ \ \qquad
\ \ \ \ \ \ \ \ \ \ \ \ \ \ \ \ \ \ \ \ \ \ \ \ \ \ \ \ \ \ \ \ \ \ \ \ \ \ \ \ \ \ \ \ \ \ \ \ \ \ \ \ \ \ \ \ \ \ \ \ \ \ \ \ \ \ }%

According to the Dirac sea \cite{DSea} concept and the Quark Model
\cite{QuarkM}, there is a u-quark sea, a d-quark sea, an s-quark sea, a
c-quark sea, a b-quark sea and so forth in the vacuum. We infer that there are
super-strong attractive (color) forces that make and hold the perfect vacuum
material. The vacuum material is a super superconductor. Its transition
temperature is much higher than the temperature at the center of the sun.
Colorless particles moving inside it look as if they are moving in completely
empty space. The super-strong forces are much stronger than the strong forces
that hold protons and neutrons inside a nucleus. The vacuum material is so
strong that atom bombs, hydrogen bombs and even the sun (a huge
continuously-exploding hydrogen bomb) cannot change it. What will happen if a
quark with one of the three colors is excited from the vacuum (the quark seas)?

\subsection{Single Quark
Excited\ \ \ \ \ \ \ \ \ \ \ \ \ \ \ \ \ \ \ \ \ \ \ \ \ \ \ \ \ \ \ \ \ \ \ \ \ \ \ \ \ \ \ \ \ \ \ \ \ \ \ \ \ \ \ \ \ \ \ \ \ \ \ \ \ \ \ \ \ \ \ \ \ \ \ \ \ \ \ \ \ \ \ \ \ \ \ \ \ \ \ \ \ \ \ \ \ \ \ \ \ \ \ \ \ \ \ \ \ \ \ \ \ \ \ \ \ \ \ \ \ \ \ \ \ \ \ \ \ \ \ \ \ \ \ \ \ \ \ \ \ \ \ \ \ \ \ \ \ \ \ \ \ \ \ \ \ \ \ \ \ \ \ \ \ \ \ \ \ \ \ \ \ \ \ \ \ \ \ \ \ \ \ \ \ \ \ \ \ \ \ \ \ \ \ \ \ \ \ \ \ \ \ \ \ \ \ \ \ \ \ \ \ \ \ \ \ \ \ \ \ \ \ \ \ \ \ \ \ \ \ \ \ \ \ \ \ \ \ \ \ \ \ \ \ \ \ \ \ \ \ \ \ \ \ \ \ \ \ \ \ \ \ \ \ \ \ \ \ \ \ \ \ \ \ \ \ \ \ \ \ \ \ \ \ \ \ \ \ \ \ \qquad
\ \ \ \ \ \ \ \ \ \ \ \ \ \ \ \ \ \ \ \ \ \ \ \ \ \ \ \ \ \ \ \ \ \ \ \ \ \ \ \ \ \ \ \ \ \ \ \ \ \ \ \ \ \ \ \ \ \ \ \ \ \ \ \ \ \ }%

It is very unlikely that three quarks (each with a different color) are
excited from the vacuum at the same time and the same location, and
immediately compose a baryon. Then they are confined inside the baryon until
an outside force breaks it. The probability is too small to explain the
confinement, because the attractive force of the quark sea is much larger than
the attractive force of a excited neighboring quark. A single quark is
constantly being excited from the vacuum, however.

If a quark q (one of u, d, s, c or b) is excited from the vacuum, its
antiquark $\overline{q}$ is born at the same time. Then, there are two
possibilities: 1) the excited quark and the excited antiquark compose a meson,
or 2) the excited quark and the excited antiquark separate into two particles.
At the same time, each of the two particles excites two accompanying excited
quarks, forming a baryon and an antibaryon.

\subsection{Accompanying Excitation
\ \ \ \ \ \ \ \ \ \ \ \ \ \ \ \ \ \ \ \ \ \ \ \ \ \ \ \ \ \ \ \ \ \ \ \ \ \ \ \ \ \ \ \ \ \ \ \ \ \ \ \ \ \ \ \ \ \ \ \ \ \ \ \ \ \ \ \ \ \ \ \ \ \ \ \ \ \ \ \ \ \ \ \ \ \ \ \ \ \ \ \ \ \ \ \ \ \ \ \ \ \ \ \ \ \ \ \ \ \ \ \ \ \ \ \ \ \ \ \ \ \ \ \ \ \ \ \ \ \ \ \ \ \ \ \ \ \ \ \ \ \ \ \ \ \ \ \ \ \ \ \ \ \ \ \ \ \ \ \ \ \ \ \ \ \ \ \ \ \ \ \ \ \ \ \ \ \ \ \ \ \ \ \ \ \ \ \ \ \ \ \ \ \ \ \ \ \ \ \ \ \ \ \ \ \ \ \ \ \ \ \ \ \ \ \ \ \ \ \ \ \ \ \ \ \ \ \ \ \ \ \ \ \ \ \ \ \ \ \ \ \ \ \ \ \ \ \ \ \ \ \ \ \ \ \ \ \ \ \ \ \ \ \ \ \ \ \ \ \ \ \ \ \ \ \ \ \ \ \ \ \ \ \ \ \ \ \ \ \ \ \ \qquad
\ \ \ \ \ \ \ \ \ \ \ \ \ \ \ \ \ \ \ \ \ \ \ \ \ \ \ \ \ \ \ \ \ \ \ \ \ \ \ \ \ \ \ \ \ \ \ \ \ \ \ \ \ \ \ \ \ \ \ \ \ \ \ \ \ \ }%

According to QCD \cite{QCD}, a quark has one of the three color charges (red,
blue or yellow), and there are strong attractive forces between quarks. Once
the quark is excited from the vacuum, it will excite other quarks in the
vacuum. Since the strong forces are saturable (a three-different-colored quark
system is a colorless one), there are only two quarks (q'$_{1}$ and q'$_{2})$
experiencing an accompanying excitation by each excited quark (q). The three
quarks have different colors. This makes the three quark system a colorless
one. Because the excited energy of q is not large enough, the quarks q'$_{1}$
and q'$_{2}$ are\ not completely excited from the vacuum. They cannot leave
their positions in the vacuum. Since the flavored quarks (s, c and b) have
large masses and flavored numbers (s, c and b), it is difficult for them to
experience accompanying excitation. The u(0)-quark and the d(0)-quark have
maximum probabilities of accompanying excitation.

Since the excited energy of q is not large enough, the two accompanying
excited quarks (q'$_{1}$ and q'$_{2}$) do not leave their positions in the
vacuum. Their color charges, electric charges and baryon numbers, however, are
temporarily excited from the vacuum state under the influence of the excited
quark q. The excited energy of the accompanying excited quarks is much smaller
than the excited energy of the excited quark q. For convenience, we estimate
that
\begin{equation}
\text{m}_{u\text{'}}\sim\text{ m}_{d\text{'}}\sim\text{ }\alpha\text{m}%
_{q},\label{Mu,d}%
\end{equation}
where M$_{q}$ is the mass of q , $\alpha$ = e$^{2}/c\hslash$ =1/137-- the fine
structure constant. The mass of a proton \cite{QuarkM}
\begin{align}
\text{M}_{\text{p}}  & \text{=}\text{ m}_{\text{q}}\text{+m}_{\text{u'}%
}\text{+ m}_{\text{d'}}\nonumber\\
& \text{=}\text{ (1+}\alpha+\alpha)\text{m}_{\text{q}}\text{ = 938.}\label{Mp}%
\end{align}
Thus from (\ref{Mu,d}) and (\ref{Mp}), we find that the u-quark inside the
proton has \
\begin{equation}
\text{B=}\frac{1}{3}\text{, S=0, s=I=}\frac{\text{1}}{\text{2}}\text{, I}%
_{z}\text{=1/2, Q=2/3, m}_{u}\text{=925 Mev;}\label{u*}%
\end{equation}
the accompanying excited u'-quark has
\begin{equation}
\text{B=}\frac{1}{3}\text{, S=0, s=I=}\frac{\text{1}}{2}\text{, I}_{z}%
\text{=}\frac{\text{1}}{2}\text{, Q=}\frac{\text{2}}{3}\text{, m}_{u^{\prime}%
}\text{=6.5 Mev;}\label{u'}%
\end{equation}
and the accompanying excited d'-quark has
\begin{equation}
\text{B=}\frac{1}{3}\text{, S=0, s=I=}\frac{\text{1}}{2}\text{, I}%
_{z}\text{=-}\frac{\text{1}}{2}\text{, Q=-1/3, m}_{d\text{'}}\text{=6.5
Mev.}\label{d'}%
\end{equation}

The accompanying excitation is temporary. When the quark $q$ is excited from
the vacuum, the two quarks (one of u'u', u'd' or d'd') undergo an accompanying
excitation due to $q$; but when $q$ leaves, the excitation
disappears.\textbf{\ }As q moves through the vacuum, it constantly locally
excites a new q'$_{1}$ / q'$_{2}$ pair, and the old q'$_{1}$ / q'$_{2}$ pair
returns to the vacuum state. Although the accompanying pair of excited quarks
is quickly changed-- one pair following another-- with the motion of $q$, two
accompanying excited quarks always appear to accompany $q$, just as an
electric field always accompanies the original electric charge. Thus not only
can we not see any individual free excited quark (q), but also we cannot see
any individual free accompanying excited quark (u' or d'). They are always
\textquotedblleft confined\textquotedblright\ inside the baryons.

Regarding the mesons, if the quark and the antiquark are separated by force,
the quark and the antiquark will excite two accompanying excited quarks for
each of them. So that a new baryon and a new antibaryon are born. We cannot
see an individual free quark here either.

To sum up, since any excited quark is always accompanied by two accompanying
excited quarks (in baryon case) or an antiquark (in meson case), an individual
free quark (q or u' or d') can never be seen. \ \ \ \ \ \ \ \ \ \ \ \ \ \ \ \ \ \ \ \ \ \ \ \ \ \ \ \ \ \ \ \ \ \ \ \ \ \ \ \ \ \ \ \ \ \ \ \ \ \ \ \ \ \ \ \ \ \ \ \ \ \ \ \ \ \ \ \ \ \ \ \ \ \ \ \ \ \ \ \ \ \ \ \ \ \ \ \ \ \ \ \ \ \ \ \ \ \ \ \ \ \ \ \ \ \ \ \ \ \ \ \ \ \ \ \ \ \ \ \ \ \ \ \ \ \ \ \ \ \ \ \ \ \ \ \ \ \ \ \ \ \ \ \ \ \ \ \ \ \ \ \ \ \ \ \ \ \ \ \ \ \ \ \ \ \ \ \ \ \ \ \ \ \ \ \ \ \ \ \ \ \ \ \ \ \ \ \ \ \ \ \ \ \ \ \ \ \ \ \ \ \ \ \ \ \ \ \ \ \ \ \ \ \ \ \ \ \ \ \ \ \ \ \ \ \ \ \ \ \ \ \ \ \ \ \ \ \ \ \ \ \ \ \ \ \ \ \ \ \ \ \ \ \ \ \ \ \ \ \ \ \ \ \ \ \ \ \ \ \ \ \ \ \ \ \ \ \ \ \ \ \ \ \ \ \ \ \ \ \ \ \ \ \ \ \ \ \ \ \ \ \ \ \ \ \ \ \ \ \ \ \ \ \ \ \ \ \ \ \ \ \ \ \ \ \ \ \ \ \ \ \ \ \ \ \ \ \ \ \ \ \ \ \ \ \ \ \ \ \ \ \ \ \ \ \ \ \ \ \ \ \ \ \ \ \ \ \ \ \ \ \ \ \ \ \ \ \ \ \ \ \ \ \ \ \ \ \ \ \ \ \ \ \ \ \ \ \ \ \ \ \ \ \ \ \ \ \ \ \ \ \ \ \ \ \ \ \ \ \ \ \ \ \ \ \ \ \ \ \ \ \ \ \ \ \ \ \ \ \ \ \ \ \ \ \ \ \ \ \ \ \ \ \ \ \ \ \ \ \ \ \ \ \ \ \ \ \ \ \ \ \ \ \ \ \ \ \ \ \ \ \ \ \ \ \ \ \ \ \ \ \ \ \ \ \ \ \ \ \ \ \ \ \ \ \ \ \ \ \ \ \ \ \ \ \ \ \ \ \ \ \ \ \ \ \ \ \ \ \ \ \ \ \ \ \ \ \ \ \ \ \ \ \ \ \ \ \ \ \ \ \ \ \ \ \ \ \ \ \ \ \ \ \ \ \ \ \ \ \ \ \ \ \ \ \ \ \ \ \ \ \ \ \ \ \ \ \ \ \ \ \ \ \ \ \ \ \ \ \ \ \ \ \ \ \ \ \ \ \ \ \ \ \ \ \ \ \ \ \ \ \ \ \ \ \ \ \ \ \ \ \ \ \ \ \ \ \ \ \ \ \ \ \ \ \ \ \ \ \ \ \ \ \ \ \ \ \ \ \ \ \ \ \ \ \ \ \ \ \ \ \ \ \ \ \ \ \ \ \ \ \ \ \ \ \ \ \ \ \ \ \ \ \ \ \ \ \ \ \ \ \ \ \ \ \ \ \ \ \ \ \ \ \ \ \ \ \ \ \ \ \ \ \ \ \ \ \ \ \ \ \ \ \ \ \ \ \ \ \ \ \ \ \ \ \ \ \ \ \ \ \ \ \ \ \ \ \ \ \ \ \ \ \ \ \ \ \ \ \ \ \ \ \ \ \ \ \ \ \ \ \ \ \ \ \ \ \ \ \ \ \ \ \ \ \ \ \ \ \ \ \ \ \ \ \ \ \ \ \ \ \ \ \ \ \ \ \ \ \ \ \ \ \ \ \ \ \ \ \ \ \ \ \ \ \ \ \ \ \ \ \ \ \ \ \ \ \ \ \ \ \ \ \ \ \ \ \ \ \ \ \ \ \ \ \ \ \ \ \ \ \ \ \ \ \ \ \ \ \ \ \ \ \ \ \ \ \ \ \ \ \ \ \ \ \ \ \ \ \ \ \ \ \ \ \ \ \ \ \ \ \ \ \ \ \ \ \ \ \ \ \ \ \ \ \ \ \ \ \ \ \ \ \ \ \ \ \ \ \ \ \ \ \ \ \ \ \ \ \ \ \ \ \ \ \ \ \ \ \ \ \ \ \ \ \ \ \ \ \ \ \ \ \ \ \ \ \ \ \ \ \ \ \ \ \ \ \ \ \ \ \ \ \ \ \ \ \ \ \ \ \ \ \ \ \ \ \ \ \ \ \ \ \ \ \ \ \ \ \ \ \ \ \ \ \ \ \ \ \ \ \ \ \ \ \ \ \ \ \ \ \ \ \ \ \ \ \ \ \ \ \ \ \ \ \ \ \ \ \ \ \ \ \ \ \ \ \ \ \ \ \ \ \ \ \ \ \ \ \ \ \ \ \ \ \ \ \ \ \ \ \ \ \ \ \ \ \ \ \ \ \ \ \ \ \ \ \ \ \ \ \ \ \ \ \ \ \ \ \ \ \ \ \ \ \ \ \ \ \ \ \ \ \ \ \ \ \ \ \ \ \ \ \ \ \ \ \ \ \ \ \ \ \ \ \ \ \ \ \ \ \ \ \ \ \ \ \ \ \ \ \ \ \ \ \ \ \ \ \ \ \ \ \ \ \ \ \ \ \ \ \ \ \ \ \ \ \ \ \ \ \ \ \ \ \ \ \ \ \ \ \ \ \ \ \ \ \ \ \ \ \ \ \ \ \ \ \ \ \ \ \ \ \ \ \ \ \ \ \ \ \ \ \ \ \ \ \ \ \ \ \ \ \ \ \ \ \ \ \ \ \ \ \ \ \ \ \ \ \ \ \ \ \ \ \ \ \ \ \ \ \ \ \ \ \ \ \ \ \ \ \ \ \ \ \ \ \ \ \ \ \ \ \ \ \ \ \ \ \ \ \ \ \ \ \ \ \ \ \ \ \ \ \ \ \ \ \ \ \ \ \ \ \ \ \ \ \ \ \ \ \ \ \ \ \ \ \ \ \ \ \ \ \ \ \ \ \ \ \ \ \ \ \ \ \ \ \ \ \ \ \ \ \ \ \ \ \ \ \ \ \ \ \ \ \ \ \ \ \ \ \ \ \ \ \ \ \ \ \ \ \ \ \ \ \ \ \ \ \ \ \ \ \ \ \ \ \ \ \ \ \ \ \ \ \ \ \ \ \ \ \ \ \ \ \ \ \ \ \ \ \ \ \ \ \ \ \ \ \ \ \ \ \ \ \ \ \ \ \ \ \ \ \ \ \ \ \ \ \ \ \ \ \ \ \ \ \ \ \ \ \ \ \ \ \ \ \ \ \ \ \ \ \ \ \ \ \ \ \ \ \ \ \ \ \ \ \ \ \ \ \ \ \ \ \ \ 

\subsection{Predictions\ and
Testing\ \ \ \ \ \ \ \ \ \ \ \ \ \ \ \ \ \ \ \ \ \ \ \ \ \ \ \ \ \ \ \ \ \ \ \ \ \ \ \ \ \ \ \ \ \ \ \ \ \ \ \ \ \ \ \ \ \ \ \ \ \ \ \ \ \ \ \ \ \ \ \ \ \ \ \ \ \ \ \ \ \ \ \ \ \ \ \ \ \ \ \ \ \ \ \ \ \ \ \ \ \ \ \ \ \ \ \ \ \ \ \ \ \ \ \ \ \ \ \ \ \ \ \ \ \ \ \ \ \ \ \ \ \ \ \ \ \ \ \ \ \ \ \ \ \ \ \ \ \ \ \ \ \ \ \ \ \ \ \ \ \ \ \ \ \ \ \ \ \ \ \ \ \ \ \ \ \ \ \ \ \ \ \ \ \ \ \ \ \ \ \ \ \ \ \ \ \ \ \ \ \ \ \ \ \ \ \ \ \ \ \ \ \ \ \ \ \ \ \ \ \ \ \ \ \ \ \ \ \ \ \ \ \ \ \ \ \ \ \ \ \ \ \ \ \ \ \ \ \ \ \ \ \ \ \ \ \ \ \ \ \ \ \ \ \ \ \ \ \qquad
\ \ \ \ \ \ \ \ \ \ \ \ \ \ \ \ \ \ \ \ \ \ \ \ \ \ \ \ \ \ \ \ \ \ \ \ \ \ \ \ \ \ \ \ \ \ \ \ \ \ \ \ \ \ \ \ \ \ \ \ \ \ \ \ \ \ }%

This proposal\ predicts:

1. There are two accompanying excited quarks inside a baryon:
\begin{equation}%
\begin{tabular}
[c]{|l|l|l|l|l|l|}\hline
{\small u}' & {\small S=C=b=0} & I=s=$\frac{1}{2}$ & {\small I}$_{z}$%
=$\frac{1}{2}$ & {\small Q=}$\frac{2}{3}${\small \ } & {\small m}$_{u^{\prime
}}${\small = 6.5,}\\\hline
{\small d}' & {\small S=C=b=0} & I=s=$\frac{1}{2}$ & {\small I}$_{z}$%
=$\frac{\text{-1}}{2}$ & {\small Q=}$\frac{\text{-1}}{3}$ & {\small m}%
$_{d^{^{\prime}}}${\small = 6.5.}\\\hline
\end{tabular}
\label{ud}%
\end{equation}
Comparing them with theoretical and experimental results of the Quark Model
\cite{QuarkM}, we find that the masses of the quarks (u and d) of the Quark
Model (m$_{u}$ = 1.5 to 4.5 Mev, m$_{\text{d}}$ = 5 to 8.5 Mev) \cite{QMass02}
are just the quantities m$_{u\text{'}}$ and m$_{d\text{'}}$, and that the
accompanying excited quarks (u'\ and d') have the same intrinsic quantum
numbers (B, S, s(spin), I, I$_{Z}$ and Q) that the quarks u and d have. Thus
the accompanying excited quark u' is the quark u \cite{QMass02} with
\begin{equation}
\text{ S=0, s=I=1/2, I}_{z}\text{=1/2, Q=2/3, m}_{u\text{'}}\text{=3
Mev;}\label{u}%
\end{equation}
and the accompanying excited quark d' is the quark d \cite{QMass02} with
\begin{equation}
\text{S=0, s=I=1/2, I}_{z}\text{=-1/2, Q=-1/3, m}_{d^{\prime}}\text{=7Mev}%
.\label{d}%
\end{equation}

2. There is always an excited quark q in a baryon. According to the Quark
Model and the accompanying excitation concept, a baryon (p or n) is composed
of three quarks (\textbf{qu'd'}). From (\ref{u}) and (\ref{d}), we get
{\small u(928) and d(930)}:
\begin{equation}%
\begin{tabular}
[c]{|l|l|l|}\hline
{\small Baryon} & {\small p(938)} & {\small n(940)}\\\hline
{\small Q. M.} & {\small uu'd'} & {\small du'd'}\\\hline
{\small m}$_{\text{u'{\small +}d'}}$ & {\small 10 Mev} & {\small 10
Mev}\\\hline
{\small q(m)} & {\small u(928)} & {\small d(930)}\\\hline
\end{tabular}
\ \ \ \text{.}\label{q*}%
\end{equation}
The q(m) has about 99\% of the mass of the baryon (p or n). This has not been
proved by experiment. There may be two reasons: 1). The mass of the q-quark is
so close to the mass of the baryon (p or n) that it is mistaken for the mass
of the baryon (p or n) in experiments. 2). Physicists have been theorizing
that the three quarks inside the baryon (p or n) have the essentially the same
masses for a long time. Since physicists have already found m$_{u\text{'}}$=3
Mev, they naturally think that another u-quark has m$_{u}=$ m$_{u\text{'}}$ =3
Mev also. Thus there is no reason to measure m$_{u} $ very carefully.

3. We have already found the electric charges of the three quarks inside a
baryon, they are not the same. We have already measured the masses of the
quarks inside SU(5) also; they are not the same either. If the three masses of
the three quarks inside a baryon are not the same, it will not break down the
SU(3) symmetries since it has more fundamental rigorous basis (see our later
paper \textquotedblleft A Proposal About the Rest Masses of
Quarks\textquotedblright, we will show that the bare masses of u, d, s, c, b,
u' and d' are essentially the same in that paper). This proposal predicts that
there is a completely excited u-quark with mass m$_{u}$ = 928 Mev inside a
proton. We suggest that experimental physicists test this prediction.

\subsection{\ Discussions\ \ \ \ \ \ \ \ \ \ \ \ \ \ \ \ \ \ \ \ \ \ \ \ \ \ \ \ \ \ \ \ \ \ \ \ \ \ \ \ \ \ }%

1. Since there really are quark seas in the vacuum and there really are
saturable strong forces between the quarks, the accompanying excitation will
definitely exist. It will help us to understand why the three quarks can
compose a stable proton.

2. Even though the form of the confinement might be different from the quarks,
there is also confinement in the leptons. We cannot see any free bare lepton,
we can see only physical leptons.

3. According to the Quark Model, a proton is composed of three
quarks.\textbf{\ }Using the masses of the three quarks (\textbf{uud}) in the
Quark Model\textbf{\ }\cite{QMass02}, we find the mass of a proton,
\begin{equation}
\text{M}_{\text{p}}\text{(QM)}=\text{m}_{u}\text{+ m}_{u}\text{+ m}_{d}%
\simeq\text{13Mev.}\label{Muud}%
\end{equation}
The experimental mass of proton M$_{\text{p}}$=938.27200$\pm$0.00004 Mev. The
three quarks (\textbf{uud) }with 13 Mev cannot make a stable proton with 938
Mev. Einstein's relativity (E =Mc$^{2})$ and the experiments of atom bombs,
hydrogen bombs and nuclear electric power stations have shown that. In fact
the proton is very stable with lifetime $\tau$
$>$
10$^{\text{25}}$\textit{years.} The theoretical mass of proton M$_{\text{p}}%
$(QM) $\simeq$13Mev is only 1.4\% of experimental value. About 98.6\% (925
Mev) of the mass has been missed. The missing mass cannot be mainly the masses
of the \textquotedblleft gluons\textquotedblright\ because the rate of the
missing mass is too large. If the gluons make up the majority of the mass of a
proton, the proton will not be so stable and the quark model would have to
change its name to the \textquotedblleft Gluon Model\textquotedblright\ (a
proton is made of gluons). As A. Pais has pointed out \cite{Confine}:
\textquotedblleft SU(3)$_{\text{C}}$ is an unbroken symmetry, i.e., the gluons
are strictly massless.\textquotedblright\ \ Thus we cannot explain the missing
mass using the gluons.

According to the accompanying excitation concept, there is, however, a
completely excited (from the vacuum) quark (u) inside the proton. The u-quark
has the missing mass (928 Mev) from (\ref{q*}). The missing mass of the proton
is the result of missing the excited u-quark (instead by another u'-quark).
Since the mass of the u-quark is so close to the mass of the proton, it is
mistaken as the mass of the proton in experiments.

4. The lack of a rigorous basis of quark confinement and the inability of the
three quarks (\textbf{uud}) of the Quark Model to compose a stable proton are
very important fundamental problems. The Standard Model is incomplete. It has
19 arbitrary parameters. This high degree of arbitrariness suggests that a
more fundamental theory underlies the Standard Model\ \cite{Standard}. The
search for solutions to these problems might find the fundamental theory.
There will have to be some new ideas that are different from the Standard
Model. In order to get the new fundamental theory, we must encourage these new ideas.

\subsection{Conclusions\ \ \ \ \ \ \ \ \ \ \ \ \ \ \ \ \ \ \ \ \ \ \ \ \ \ \ \ \ \ \ \ \ \ \ \ \ \ \ \ \ \ \ \ \ \ \ \ \ \ \ \ \ \ \ \ \ \ \ \ \ \ \ \ \ \ \ \ \ \ \ \ \ \ \ \ \ \ \ \ \ \ \ \ \ \ \ \ \ \ \ \ \ \ \ \ \ \ \ \ \ \ \ \ \ \ \ \ \ \ \ \ \ \ \ \ \ \ \ \ \ \ \ \ \ \ \ \ \ \ \ \ \ \ \ \ \ \ \ \ \ \ \ \ \ \ \ \ \ \ \ \ \ \ \ \ \ \ \ \ \ \ \ \ \ \ \ \ \ \ \ \ \ \ \ \ \ \ \ \ \ \ \ \ \ \ \ \ \ \ \ \ \ \ \ \ \ \ \ \ \ \ \ \ \ \ \ \ \ \ \ \ \ \ \ \ \ \ \ \ \ \ \ \ \ \ \ \ \ \ \ \ \ \ \ \ \ \ \ \ \ \ \ \ \ \ \ \ \ \ \ \ \ \ \ \ \ \ \ \ \ \ \ \ \ \ \ \ \ \ \ \ \ \ \ \ \ \ \ \ \ \ \ \ \ \ \ \ \ \ \ \ \qquad
\ \ \ \ \ \ \ \ \ \ \ \ \ \ \ \ \ \ \ \ \ \ \ \ \ \ \ \ \ \ \ \ \ \ \ \ \ \ \ \ \ \ \ \ \ \ \ \ \ \ \ \ \ \ \ \ \ \ \ \ \ \ \ \ \ \ \ \ \ \ \ \ \ \ \ \ \ \ \ \ \ \ \ \ \ \ \ \ \ \ \ \ \ \ \ \ \ \ \ }%

1. There are two accompanying excited quarks inside a baryon.

2. Because any excited quark is always accompanied by two accompanying excited
quarks (in baryon case) or an antiquark (in meson case), an individual free
quark can never be seen.\ Therefore the quarks are always \textquotedblleft
confined\textquotedblright\ inside the hadrons .

\ \ \ \ \ \ \ \ \ \ \ \ \ \ \ \ \ \ \ \ \ \ \ \ \ \ \ \ \ \ \ \ \ \ \ \ \ \ \ \ \ \ \ \ \ \ \ \ \ \ \ \ \ \ \ \ \ \ \ \ \ \ \ \ \ \ \ \ \ \ \ \ \ \ \ \ \ \ \ \ \ \ \ \ \ \ \ \ \ \ \ \ \ \ \ \ \ \ \ \ \ \ \ \ \ \ \ \ \ \ \ \ \ \ \ \ \ \ \ \ \ \ \ \ \ \ \ \ \ \ \ \ \ \ \ \ \ \ \ \ \ \ \ \ \ \ \ \ \ \ \ \ \ \ \ \ \ \ \ \ \ \ \ \ \ \ \ \ \ \ \ \ \ \ \ \ \ \ \ \ \ \ \ \ \ \ \ \ \ \ \ \ \ \ \ \ \ \ \ \ \ \ \ \ \ \ \ \ \ \ \ \ \ \ \ \ \ \ \ \ \ \ \ \ \ \ \ \ \ \ \ \ \ \ \ \ \ \ \ \ \ \ \ \ \ \ \ \ \ \ \ \ \ \ \ \ \ \ \ \ \ \ \ \ \ \ \ \ \ \ \ \ \ \ \ \ \ \ \ \ \ \ \ \ \ \ \ \ \ \ \ \ \ \ \ \ \ \ \ \ \ \ \ \ \ \ \ \ \ \ \ \ \ \ \ \ \ \ \ \ \ \ \ \ \ \ \ \ \ \ \ \ \ \ \ \ \ \ \ \ \ \ \ \ \ \ \ \ \ \ \ \ \ \ \ \ \ \ \ \ \ \ \ \ \ \ \ \ \ \ \ \ \ \ \ \ \ \ \ \ \ \ \ \ \ \ \ \ \ \ \ \ \ \ \ \ \ \ \ \ \ \ \ \ \ \ \ \ \ \ \ \ \ \ \ \ \ \ \ \ \ \ \ \ \ \ \ \ \ \ \ \ \ \ \ \ \ \ \ \ \ \ \ \ \ \ \ \ \ \ \ \ \ \ \ \ \ \ \ \ \ \ \ \ \ \ \ \ \ \ \ \ \ \ \ \ \ \ \ \ \ \ \ \ \ \ \ \ \ \ \ \ \ \ \ \ \ \ \ \ \ \ \ \ \ \ \ \ \ \ \ \ \ \ \ \ \ \ \ \ \ \ \ \ \ \ \ \ \ \ \ \ \ \ \ \ \ \ \ \ \ \ \ \ \ \ \ \ \ \ \ \ \ \ \ \ \ \ \ \ \ \ \ \ \ \ \ \ \ \ \ \ \ \ \ \ \ \ \ \ \ \ \ \ \ \ \ \ \ \ \ \ \ \ \ \ \ \ \ \ \ \ \ \ \ \ \ \ \ \ \ \ \ \ \ \ \ \ \ \ \ \ \ \ \ \ \ \ \ \ \ \ \ \ \ \ \ \ \ \ \ \ \ \ \ \ \ \ \ \ \ \ \ \ \ \ \ \ \ \ \ \ \ \ \ \ \ \ \ \ \ \ \ \ \ \ \ \ \ \ \ \ \ \ \ \ \ \ \ \ \ \ \ \ \ \ \ \ \ \ \ \ \ \ \ \ \ \ \ \ \ \ \ \ \ \ \ \ \ \ \ \ \ \ \ \ \ \ \ \ \ \ \ \ \ \ \ \ \ \ \ \ \ \ \ \ \ \ \ \ \ \ \ \ \ \ \ \ \ \ \ \ \ \ \ \ \ \ \ \ \ \ \ \ \ \ \ \ \ \ \ \ \ \ \ \ \ \ \ \ \ \ \ \ \ \ \ \ \ \ \ \ \ \ \ \ \ \ \ \ \ \ \ \ \ \ \ \ \ \ \ \ \ \ \ \ \ \ \ \ \ \ \ \ \ \ \ \ \ \ \ \ \ \ \ \ \ \ \ \ \ \ \ \ \ \ \ \ \ \ \ \ \ \ \ \ \ \ \ \ \ \ \ \ \ \ \ \ \ \ \ \ \ \ \ \ \ \ \ \ \ \ \ \ \ \ \ \ \ \ \ \ \ \ \ \ \ \ \ \ \ \ \ \ \ \ \ \ \ \ \ \ \ \ \ \ \ \ \ \ \ \ \ \ \ \ \ \ \ \ \ \ \ \ \ \ \ \ \ \ \ \ \ \ \ \ \ \ \ \ \ \ \ \ \ \ \ \ \ \ \ \ \ \ \ \ \ \ \ \ \ \ \ \ \ \ \ \ \ \ \ \ \ \ \ \ \ \ \ \ \ \ \ \ \ \ \ \ \ \ \ \ \ \ \ \ \ \ \ \ \ \ \ \ \ \ \ \ \ \ \ \ \ \ \ \ \ \ \ \ \ \ \ \ \ \ \ \ \ \ \ \ \ \ \ \ \ \ \ \ \ \ \ \ \ \ \ \ \ \ \ \ \ \ \ \ \ \ \ \ \ \ \ \ \ \ \ \ \ \ \ \ \ \ \ \ \ \ \ \ \ \ \ \ \ \ \ \ \ \ \ \ \ \ \ \ \ \ \ \ \ \ \ \ \ \ \ \ \ 

\begin{center}
\bigskip\textbf{Acknowledgment}
\end{center}

I sincerely thank Professor Robert L. Anderson for his valuable advice. I
thank Professor D. P. Landau very much for his help. I thank Dr. Xin Yu very
much for his help also.

\end{document}